\documentclass{ar-1col-S2O}
\usepackage[numbers]{natbib}
\usepackage{url}
\usepackage{xspace}
\usepackage[unicode=true,
   linktocpage,
   linkbordercolor={0.5 0.5 1},
  citebordercolor={0.5 1 0.5},
  linkcolor=blue]{hyperref}
\usepackage{xcolor}%
\usepackage{bm}
\usepackage{dutchcal}
\usepackage{amsmath}
\usepackage{amssymb}
\usepackage{booktabs}
\usepackage{array}

\newcommand{\sro}{Sr\ensuremath{_2}RuO\ensuremath{_4}\xspace}
\newcommand{\pxpy}{\ensuremath{p_x\pm i p_y}\xspace}
\newcommand{\ute}{UTe\ensuremath{_2}\xspace}
\newcommand{\upt}{UPt\ensuremath{_3}\xspace}
\newcommand{\bkfa}{Ba\ensuremath{_{1-x}}K\ensuremath{_x}Fe\ensuremath{_2}As\ensuremath{_2}\xspace}
\newcommand{\Tc}{\ensuremath{T_{\rm c}}\xspace}

\jname{Xxxx. Xxx. Xxx. Xxx.}
\jvol{AA}
\jyear{YYYY}
\doi{10.1146/((please add article doi))}

\begin{document}

\markboth{Ramshaw}{Symmetry, Sound, and the Quest for Chiral Superconductors}

\title{Symmetry, Sound, and the Quest for Chiral Superconductors}

\author{B. J. Ramshaw$^1$ 
\affil{$^1$Laboratory of Atomic and Solid State Physics, Cornell University, Ithaca, NY, USA, 14853; email: bradramshaw@cornell.edu}}

\begin{abstract}
The confirmation of a bulk, topological superconductor is a central goal of modern quantum materials research. Despite decades of searching, no confirmed examples yet exist. A primary difficulty is that most experimental signatures of topological superconductivity are ambiguous and indirect. I review efforts to narrow the search for a chiral topological superconductor using measurements of elastic moduli. In principle, these experiments can provide unambiguous thermodynamic evidence for multi-component superconductivity---a pre-requisite for chiral superconductivity in 2D and quasi-2D materials. Despite the potential strength of this program, the results are still negative. I provide a pedagogical introduction to multi-component order parameters, chiral topological superconductivity, and how superconducting order parameters couple to strain. I discuss ongoing efforts in heavy fermion, iron pnictide, and Kagome superconductors, and suggest routes for speeding up the search process, as well as other experimental tools that should be developed for the pursuit of this goal.
\end{abstract}

\maketitle

\tableofcontents

\section{INTRODUCTION}
Topological superconductivity has entered the popular physics lexicon because of its possible applications to quantum computing \cite{kitaevFaulttolerantQuantumComputation2003}. A distinct lack of verifiable topological superconductors has impeded this application. The central difficulty is the absence of experiments that can definitively answer the question ``is this superconductor topological?'' While the short answer is ``there is no single such experiment", this review outlines how ultrasound---measuring the strain susceptibility of a superconducting order parameter---can play a role in answering this question. 

At a broad level, this review is really about multi-component order parameters. These are order parameters that are necessarily described by more than one number. A simple example is an in-plane ferromagnetic moment in a tetragonal crystal: both $x$ and $y$ components of the magnetization $\bf{M}$ are required for a complete description of the ordered state, making the order parameter a two-dimensional (axial) vector, ${\bf M}=\left\{M_x, M_y\right\}$. While the multi-component nature of magnetization is obvious, it is less obvious that this situation can apply to superconductivity, which is typically described by a single (complex) order parameter. Under the right circumstances, however, multi-component superconducting order parameters are favorable; under an even more restrictive set of circumstances, they can be chiral; more restrictive still, they can be topological \cite{satoTopologicalSuperconductorsReview2017}.

Multi-component superconducting order parameters have unique degrees of freedom---the relative phase and amplitude of the different components---that can directly couple to shear strain in a way that the degrees of freedom for a single-component superconductor cannot. I focus on a specific thermodynamic quantity---the elastic modulus tensor---that can access these relative degrees of freedom and can therefore indicate whether a superconductor has a multi-component order parameter. In short, finding a singular drop---what I will refer to as a ``jump"---in a shear elastic modulus at \Tc provides direct, thermodynamic evidence for a multi-component superconducting state, and one that could be topological. While identifying only an ``ingredient'' for topology might seem underwhelming, the lack of direct experimental probes for topological invariants---and the lack of topological superconductors---makes this nonetheless an attractive program. 

This review is organized as follows: \autoref{se:toposc} provides an overview of chiral superconductivity and describes in which sense it can be topological; \autoref{se:ultra} describes how strains couple to superconducting order parameters, for both single- and multi-component superconductors, and the experimental consequences of this coupling; \autoref{se:experiments} describes the current state of the field, including work on \sro, \upt, \bkfa, and \ute; and \autoref{se:future} outlines some promising directions, including hexagonal point groups, 2D materials, and new, related experimental techniques. 

I focus on the experimental phenomenology of chiral superconductors, and do not attempt to give an exhaustive treatment of topological superconductivity (either experimental or theoretical). For that, the interested reader should refer to Sato and Ando \cite{satoTopologicalSuperconductorsReview2017}. I also do not discuss ultrasonic \textit{attenuation} measurements on superconductors: while powerful and of great historical significance for the field, these are transport rather than thermodynamic experiments and therefore are less straightforward to interpret. The definitions provided in the margins, e.g. of an ``irreducible representation", should be understood in the context of symmetry-breaking phase transitions and not as rigorous mathematical definitions.

\section{MULTI-COMPONENT, CHIRAL, AND TOPOLOGICAL SUPERCONDUCTIVITY}
\label{se:toposc}

The subset of topological superconductors discussed in this review are those formed from a multi-component order parameter that has a non-trivial winding number in 2D momentum space \cite{satoTopologicalSuperconductorsReview2017}. While at first this might seem highly restrictive, many notable experimental claims of topological superconductivity fall into this category, including the classic $p_x + i p_y$ state. I will refer to these as chiral topological superconductors. 

\subsection{Multi-component superconductors}

\begin{textbox}[b]
	\section{Accidentally-Degenerate Order Parameters}
	Any two order parameters that condense at finite temperature, but that are not related by symmetry, are said to be ``accidentally" two component. An example is an admixture of $s$ into a $d_{x^2-y^2}$ state to form a $d_{x^2-y^2}+is$ state. Accidental two-component order parameters have been proposed for \sro \cite{ramiresSuperconductingOrderParameter2019,kivelsonProposalReconcilingDiverse2020}, \upt \cite{saulsOrderParameterSuperconducting1994}, and \ute \cite{hayesMulticomponentSuperconductingOrder2021}, and are thus more relevant than one might initially suppose. Because the two order parameters are not related by symmetry, there is no reason that they need to condense at the same transition temperature and generically there will be two \Tc's, where time reversal symmetry (TRS) may be broken at the lower transition once the second component appears. 
\end{textbox}

At a minimum, chiral superconductivity requires something that can \textit{wind}. More specifically, we need to define a vector associated with the order parameter that can form a structure with a non-trivial skyrmion number in the Brillouin zone \cite{toulousePrinciplesClassificationDefects1976}. This requires a ``multi-component" superconducting order parameter---an order parameter that forms a multi-dimensional irreducible representation (irrep.) of the point group (see sidebar for the case of accidentally-degenerate single component order parameters).

\begin{marginnote}[]
	\entry{Irreducible representation}{A set of objects, such as order parameter components or strains, that mixes only with itself under a set of symmetry operations.}
	\entry{Order parameter}{A quantity that characterizes a symmetry-breaking phase transition. It is zero in the high-symmetry unordered phase and nonzero in the symmetry-broken ordered phase.}
	\entry{Multi-component}{An object that transforms as a two (or higher) dimensional irrep. of the point group, e.g. magnetization in a tetragonal crystal.}
\end{marginnote}

Multi-component superconducting order parameters can occur when the lattice symmetry is sufficiently high such that there are degenerate configurations for the orbital part of the Cooper pair wavefunction. The $p_x + ip_y$ state on a lattice with at least $C_4$ rotational symmetry is an example of this: if $x$ and $y$ are symmetric, then both $p_x$ and $p_y$ components must be considered equally. The full superconducting order parameter is then described by two complex numbers---two superconducting amplitudes and two phases---that we write as ${\bf \Psi}=\left\{\Delta_x e^{i\phi_x},\Delta_ye^{i\phi_y}\right\}$. The shorthand $\left\{p_x, p_y\right\}$ is really an indication that the order parameter is formed from an $l = 1$ irrep. of the point group: in a tetragonal system, this is the $E_u$ irrep. The shorthand is then a stand-in for ${\bf \Psi}=\left\{\Delta_x\sin k_x e^{i\phi_x},\Delta_y\sin k_y e^{i\phi_y}\right\}$, where the $\Delta_i$ now transform trivially under the point group and the $\sin k_i$ terms enforce the $p$-wave symmetry of the order parameter. I will use the shorthand notation when convenient, and the more explicit form when necessary.

While $p$-wave is the most discussed multi-component state, it is not the only one. In \upt, for example, the ground state is thought to be the two-component $\left\{f_{z(x^2-y^2)},f_{xyz}\right\}$, or $E_{2u}$, state \cite{saulsOrderParameterSuperconducting1994}. Multi-component order parameters are not restricted to odd-parity order parameters: the two-component, even-parity $\left\{d_{xz},d_{yz}\right\}$ ($E_g$) state has been proposed for \sro \cite{benhabibUltrasoundEvidenceTwocomponent2021}, and the even-parity $\left\{d_{x^2-y^2},d_{xy}\right\}$ ($E_{2g}$), state is a possibility in hexagonal systems. 

\begin{marginnote}[]
	\entry{Odd parity}{An object that is odd (i.e. changes sign) under the inversion operation, such as a polar vector (e.g. an electric field).}
	\entry{Even parity}{An object that is even under the inversion operation, such as an axial vector (e.g. a magnetic field).}
	\entry{Chiral}{An object that possesses definite handedness, such that it is distinct from its mirror image. }
\end{marginnote}


\subsection{Chiral superconductors}

At \Tc, the superconductor spontaneously breaks the lattice (and gauge) symmetry and one of these components---or a superposition of them---orders. For example, the two-component $\left\{p_x, p_y\right\}$ order parameter can order as $p_x$ or $p_y$ (nematic), $p_x \pm p_y$ (diagonal nematic), or $p_x \pm i p_y$ (chiral). Which of these 6 states is the ground state depends on details of the pairing interaction and on the electronic structure. 

Under the right circumstances, the chiral $\Psi_1 \pm i \Psi_2$ state will be the ground state. Again using $p$-wave as an illustrative example, it is often argued that the chiral $p_x \pm i p_y$ state will be energetically favoured because it opens a full gap on the Fermi surface \cite{riceSr2RuO4ElectronicAnalogue1995}: the nematic $p_x$ state, for example, has a $\sin k_x$-like gap with nodes at $k_x =0$, whereas the $p_x \pm i p_y$ state has a full gap that transforms as $\sqrt{\sin^2 k_x + \sin^2 k_y}$. Whether or not the chiral state actually has larger condensation energy than the other states again depends on details, but it is plausible that the chiral state can win out.

As emphasized above, triplet versus singlet, and odd versus even parity, is no constraint up to this point: $d_{xz}\pm i d_{yz}$ is also a chiral superconductor. Because the chiral state necessarily breaks time reversal symmetry---$\Psi_1 + i \Psi_2$ transforms to $\Psi_1 - i \Psi_2$ under time reversal---these states permit chiral edge currents by symmetry. Does that make them topological, and are these currents charge-neutral Majorana edge modes? Not necessarily: the answer depends on details. 

\subsection{Chiral topological superconductors}

I will illustrate how a multi-component order parameter can lead to a chiral topological state using the $p_x \pm ip_y$ state in a tetragonal lattice, and on a single quasi-2D fermi, surface as an example. 

To begin, it is helpful to write the Bogoliubov-de Gennes Hamiltonian in the following notation \cite{raghuHiddenQuasiOneDimensionalSuperconductivity2010}:
\begin{equation}
	H_{\rm BdG} = \sum_{\bf k}\Psi^{\dagger}_{\bf k}\left[\left({\vec \delta}({\bf k}) \cdot{\vec \tau}\right)\otimes \mathbb{I}\right]\Psi_{\bf k},
	\label{eq:bdg}
\end{equation}
where the Nambu spinors are defined as $\Psi_{\bf k}^{\dagger}=\left\{c_{\bf k \uparrow}^\dagger, c_{\bf k \downarrow}^\dagger, c_{\bf -k \downarrow}, -c_{\bf -k \uparrow} \right\}$, $\vec\tau$ is the vector of Pauli matrices in particle-hole space, $\mathbb{I}$ is the $2\times2$ identity matrix, and $\vec \delta({\bf k})$ is the pseudo-Zeeman field given by 
\begin{equation}
	\vec \delta({\bf k}) = \left\{\Re\left[\Delta({\bf k})\right],\Im\left[\Delta({\bf k})\right],\xi({\bf k})\right\},
	\label{eq:zeeman}
\end{equation}
where $\Delta({\bf k})$ is the superconducting gap function in the ordered state and $\xi({\bf k})$ is the single-particle energy measured from the chemical potential. Where did the ``multi-component" nature of the gap go? At \Tc, the order parameter condenses into a single component (or a single superposition of components) and breaks the symmetry of the high-temperature phase.

\begin{figure}[ht!]
	\centering
	\includegraphics[width=\textwidth]{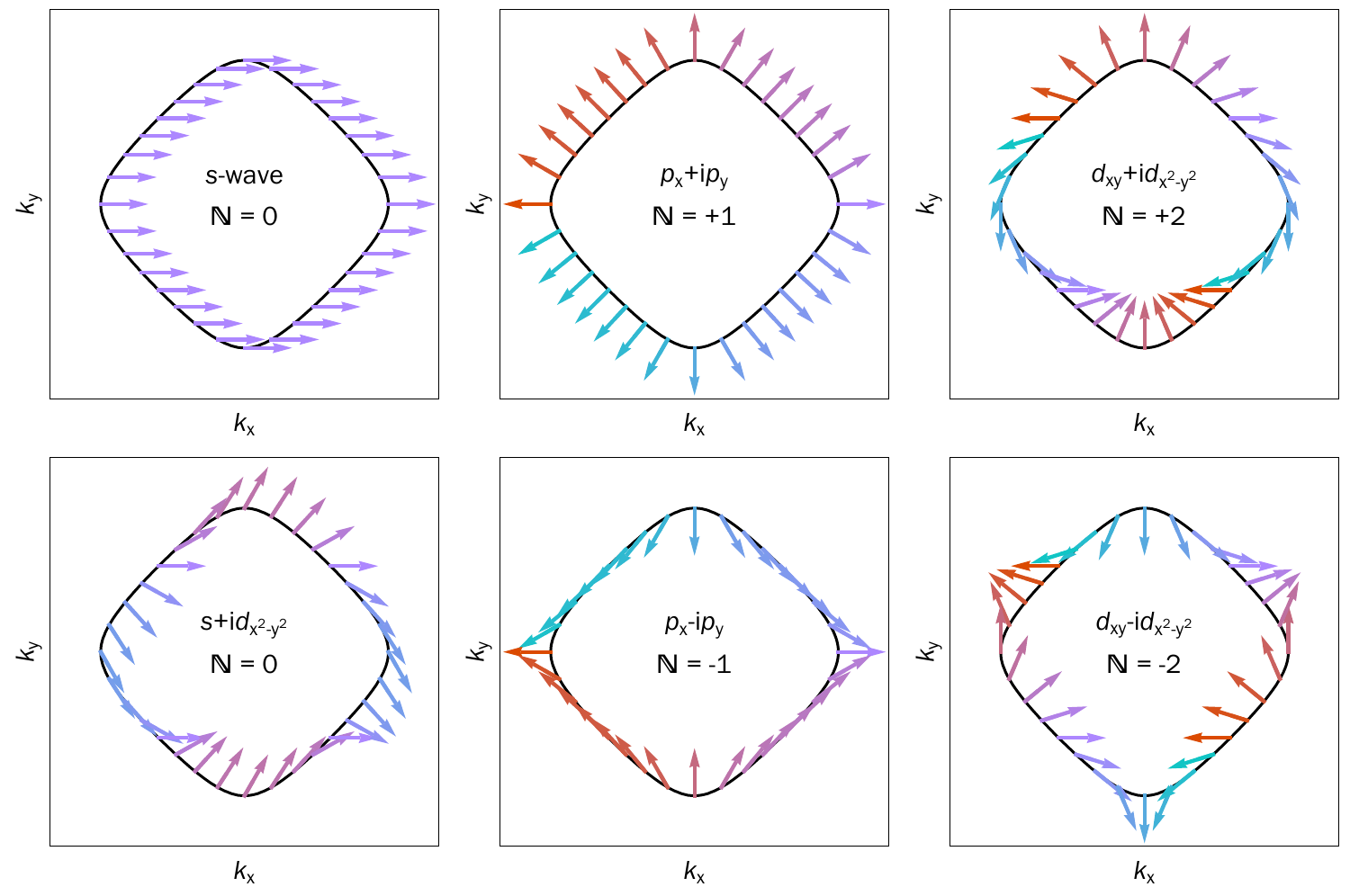}
	\caption{\textbf{Skyrmion numbers for several gaps.} The 6 panels show fermi surfaces (solid black lines) with different superconducting gap functions labeled in the center. Arrow directions (and, equivalently, their colors) indicate the direction of the pseudo-Zeeman field given by \autoref{eq:zeeman} (note that this quantity is defined everywhere in the Brillouin zone but only show on the fermi surface for clarity). The $s$-wave state is topologically trivial, with skyrmion number 0. The $d+is$ state breaks time reversal symmetry, but its pseudo-Zeeman field is trivially related to the $s$-wave case by a smooth deformation (hence the state is not chiral despite breaking TRS). The \pxpy and $d_{xy}\pm id_{x^2-y^2}$ cases, on the other hand, are topologically distinct---there is no smooth transformation from the state with $\mathbb{N}=+1$ to the state with $\mathbb{N}=-1$, nor to the $\mathbb{N}=0$ state (note that the $d_{xy}\pm id_{x^2-y^2}$ is shown in a $C_4$ symmetric Brillouin zone for clarity).  }
	\label{fig:skyrmion}
\end{figure}

The vector defined by \autoref{eq:zeeman} is the key element here: this is the object that can have a non-trivial topological number in the 2D Brillouin zone (see \autoref{fig:skyrmion}). Specifically, the topology is quantified by the skyrmion number 
\begin{equation}
	\mathbb{N}=\frac{1}{4\pi}\int_{\rm BZ}d^2k~ \hat{\delta}({\bf k})\cdot \left( \frac{\partial}{\partial k_x}\hat{\delta}({\bf k})\times \frac{\partial}{\partial k_y}\hat{\delta}({\bf k})\right), 
	\label{eq:skyrmion}
\end{equation}
where $\hat{\delta}({\bf k}) \equiv \frac{\vec{\delta}({\bf k})}{\left|\vec{\delta}({\bf k})\right|}$. The state is topological when $\mathbb{N}\neq0$, with $\left|\mathbb{N}\right|$ chiral protected edge modes. Clearly $\mathbb{N}$ is not defined where the gap goes to zero, hence why nodal states are not topological in this sense\footnote{Nodal states can still have topological numbers defined on lower-dimensional spaces, such as a closed path around a node, in analogy with Weyl and Dirac semimetals.}. 

The skyrmion number $\mathbb{N}$ determines the number of Majorana fermion modes at the sample boundaries/domain walls and in vortex cores. If $\mathbb{N}=1$, as is the case for \pxpy on a single 2D fermi surface, then the mode in the vortex core is pinned to zero energy, while the edge mode disperses linearly. Note that I have dealt with the case of a single band; if there are multiple bands, $\mathbb{N}$ should be summed over all bands, possibly leading to a topologically trivial state even if $\mathbb{N} \neq 0$ on any particular band. The chiral edge currents will still be present in principle, but are not topologically protected and can be localized by disorder, which scatters pairs from one Fermi surface to another \cite{raghuHiddenQuasiOneDimensionalSuperconductivity2010}.

Again, I have not directly invoked singlet versus triplet pairing---topological singlet states are possible. For example, the $d_{x^2-y^2}$ state is degenerate with $d_{xy}$ state in hexagonal and rhombohedral systems. This can form a topological $d_{x^2-y^2}\pm id_{xy}$ state, with $\mathbb{N}=\pm 2$. Such a state has been predicted to emerge from proximitizing two known $d_{x^2-y^2}$ superconductors rotated 45 degrees with respect to one another \cite{canHightemperatureTopologicalSuperconductivity2021}. This provides the motivation behind such efforts in Bi$_2$Sr$_2$CaCu$_2$O$_{8+x}$  \cite{zhaoTimereversalSymmetryBreaking2023}. It is worth noting that $\mathbb{N}=\pm 2$ produces two Majorana modes in each vortex core that, generically, will hybridize and no longer sit at zero energy, making them unsuitable for particular applications. It is in this sense that triplet pairing is ``desirable"---in the limit of no spin orbit coupling and a centrosymmetric crystal structure, a triplet spin configuration means an odd-parity orbital configuration with an odd number of topologically protected zero modes. 

The connection between time reversal symmetry breaking and topology is clear from the form of \autoref{eq:zeeman}: in order for $\vec{\delta}({\bf k})$ to have a non-trivial structure in the Brillouin zone, it needs to have both real and imaginary components (otherwise the vector always points in the same direction). \autoref{fig:skyrmion} visualizes $\vec{\delta}({\bf k})$ around the Fermi surface for the $s$, $d_{x^2-y^2}+is$, and $p_x \pm i p_y$ states (note that $\vec{\delta}({\bf k})$ is defined over the entire zone, but I show it only on the FS for clarity). As illustrated by the $d_{x^2-y^2}+is$ example, it is also clear that time reversal symmetry breaking is a necessary but not sufficient condition for topology: the $d_{x^2-y^2}+is$ state has a trivial skyrmion number, even though it breaks time reversal symmetry and has a full gap.

\section{SUPERCONDUCTIVITY AND ULTRASOUND }
\label{se:ultra}
\begin{marginnote}[]
	\entry{Elastic Modulus}{Often referred to as an elastic constant, an elastic modulus $c_{ijkl}$ relates the stress $\sigma_{ij}$ to the strain $\varepsilon_{kl}$ through Hooke's law: $\sigma_{ij} = c_{ijlk}\varepsilon_{kl}$.   }
	\entry{Sound Velocity}{One way to measure an elastic modulus is through its associated speed of sound, $v_{ijlk} =\sqrt{\frac{c_{ijkl}}{\rho}}$, where $\rho$ is the material density. Voigt notation is typically used such that $v_{xxxx}$ is written $v_{11}$, for example. }
	\entry{Ultrasound}{High-frequency ($>20$ kHz) acoustic waves, typically generated using a piezoelectric transducer attached to (or grown  onto) the sample under study. }
\end{marginnote}

Ultrasound measurements have played a central role in the study of superconductors for more than 80 years \cite{bardeenTheorySuperconductivity1957,morseSuperconductingEnergyGap1957}. Like the specific heat, ultrasound accesses a thermodynamic quantity: for a free energy density $f$, the specific heat is $C=-\frac{1}{T}\frac{\partial^2f}{\partial T^2}$, and the elastic moduli are $c_{ijkl}=\frac{\partial^2f}{\partial \varepsilon_{ij}\partial\varepsilon_{kl}}$, where $\varepsilon_{ij}$ are elements of the strain tensor. Because the strain tensor contains up to 6 distinct elements (depending on the crystal symmetry), the response of the elastic tensor---or, equivalently, the sound velocity as a function of direction and polarization---to superconductivity contains symmetry information about the superconducting order parameter. And unlike magnetic susceptibility, it remains a bulk probe even in the superconducting state because sound waves are carried by the inertia of the lattice, which is unscreened.

An obvious question is why superconductivity should care about the elastic stiffness of the lattice\footnote{Here I am referring to superconductivity in general, independent of whether or not phonons are involved in the pairing mechanism.}. The short answer is because band structures depend on atomic coordinates, and sound waves distort those coordinates. Put another way, if the electronic structure knows about the lattice---i.e. if it is not Galilean invariant---then electronic properties are sensitive to distortions of the lattice.

To estimate the strain sensitivity of the electronic structure, consider a simple tight binding model on a square lattice, with lattice constant $a$. Placing $s$ orbitals of characteristic size $r_0$ on each site, the strain $\varepsilon_{xx}$ modifies the nearest-neighbor hopping $t$ to first order in strain by $t\rightarrow t(1+\alpha_{xx} \varepsilon_{xx})$, where $\alpha_{xx} \equiv \frac{1}{1-2 r_0/a}$. Because lattice spacings are of order twice the size of the characteristic orbital size---$a \approx 2 r_0$---the coefficient $\alpha_{xx}$ can be large. In \sro, for example, it is experimentally determined that $\alpha_{xx} \approx 7$ \cite{barberRoleCorrelationsDetermining2019,liHighsensitivityHeatcapacityMeasurements2021}. Similar analysis shows that shear strains will have a similar-sized effect on the next-nearest-neighbor hopping, $t'$, and the effect is larger for orbitals with higher angular momentum\footnote{The hopping integral between two orbitals scales with distance like $d^{(l+l'+1)}$, where $l$ and $l'$ are the orbital angular momentum quantum numbers of the two orbitals \cite{harrisonUniversalLinearcombinationofatomicorbitalsParameters1980}, and hence $\alpha \approx (l+l'+1)$.}.

Clearly, strains have an $\mathcal{O}(1)$ effect on the electronic structure. At a minimum, this will affect superconductivity through the density of states (DOS). It can also modulate the pairing potential directly, for example through modulation of exchange interactions.

\begin{figure}[ht!]
	\centering
	\includegraphics[width=0.8\textwidth]{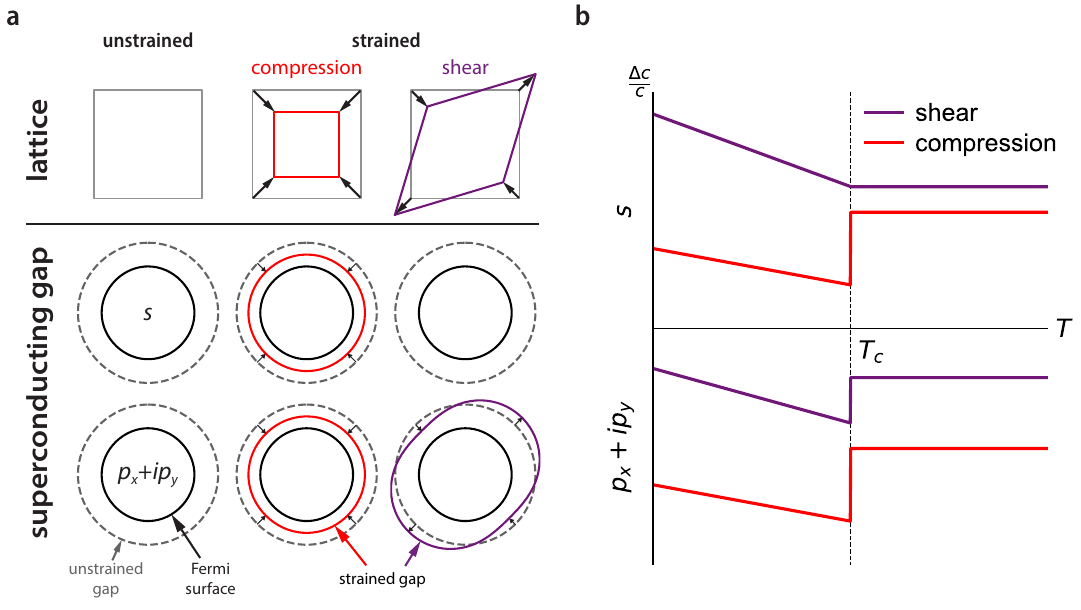}
	\caption{\textbf{Jumps for single and two-component order parameters.} \textbf{a}, An illustration of how two example order parameters---single-component $s$-wave and two-component \pxpy---respond to strain. Both gaps respond to scalar strain, while only the two-component gap couples to shear strain. \textbf{b}, The qualitative changes in elastic moduli across \Tc for one- and two-component order parameters. All superconductors have a discontinuity in their scalar moduli across \Tc, but only two-component superconductors have discontinuities in their shear moduli. From \cite{theussSinglecomponentSuperconductivityUTe22024a}.  }
	\label{fig:jumps}
\end{figure}

Putting aside the microscopic mechanisms of coupling between strain and superconductivity, it is helpful to consider these couplings purely on symmetry grounds. This is where the power of ultrasound becomes apparent. Strains fall into two general classes: ``volumetric'' or ``scalar'' strains, which preserve the symmetries of the lattice but change the volume of the unit cell; and ``shear'' strains, which break the lattice symmetry but preserve its volume (to first order in strain, see \autoref{fig:jumps})\footnote{The words ``longitudinal'' and ``transverse'' are often used in the context of pulse echo ultrasound. These refer to whether the polarization vector of the strain field is parallel (longitudinal) or or perpendicular (transverse) to the strain wave propagation vector. Whether or not these correspond to irreducible strains depends on the crystal symmetry and the propagation and polarization directions.}.

Schematically, because scalar strains change volumes, they couple directly to the magnitude of the superconducting gap. This coupling produces a discontinuity at \Tc that mirrors the discontinuity measured in the specific heat. Shear strains have only higher-order couplings and therefore generally lack a discontinuity at \Tc. However, if the superconducting order parameter is multi-component, shear strain can couple to the relative phase and amplitudes of the two order parameter components, and this coupling will then produce a discontinuity at \Tc.

At the simplest level, the program of searching for a potential chiral topological superconductor with ultrasound is simply one of looking for jumps in shear elastic moduli at \Tc. If one is found, then at the very least there is a multi-component superconducting state, irrespective of the microscopic details of the superconductivity. Of course, nothing is ever this simple. 

\subsection{Strain and Elastic Moduli}

Elastic moduli are thermodynamic coefficients---they are the second derivatives of the free energy with respect to strain. Coupled with the fact that bulk, three-dimensional superconductors are usually well-described by mean field theory, this means that clear, symmetry-based statements can be made about superconducting order parameters by analyzing the changes in their elastic moduli at \Tc. 

It is most straightforward to work with the (irreps.) of strain for the particular point group in question. In a $D_{\rm 4h}$ tetragonal system, for example, the unique strains are $\varepsilon_{A_{1g}^1} = \varepsilon_{xx}+\varepsilon_{yy}$,  $\varepsilon_{A_{1g}^2} = \varepsilon_{zz}$, $\varepsilon_{B_{1g}} = \varepsilon_{xx}-\varepsilon_{yy}$, $\varepsilon_{B_{2g}} = \varepsilon_{xy}$, and $\varepsilon_{E_g} = \left\{\varepsilon_{xz},\varepsilon_{yz}\right\}$. This labeling will make the coupling between strain and order parameters transparent and avoid clutter.

I start by defining the elastic free energy density of a solid as 
\begin{equation}
	f_{\rm el} = \frac{1}{2}\sum_{\Gamma}c_{\Gamma,\Gamma}\varepsilon_{\Gamma}^2 + \sum_{\Gamma\neq \Gamma'}c_{\Gamma,\Gamma'}\varepsilon_{\Gamma}\varepsilon_{\Gamma'},
	\label{eq:fel}
\end{equation}
where $\Gamma$ labels the irreducible representations of strain. The first term is Hooke's law, and it is conventional to write $c_{\Gamma}\equiv c_{\Gamma,\Gamma}$ for these diagonal terms. The second term contains the cross-couplings between different strains of the same representation, i.e. the coupling between in-plane $\varepsilon_{xx}+\varepsilon_{yy}$ compression with out-of-plane $\epsilon_{zz}$ expansion---when you squish something in one direction, it expands in another.

The elastic moduli are defined as the second derivatives of this free energy density with respect to strain:
\begin{equation}
	c_{\Gamma,\Gamma'} \equiv \frac{\partial^2 f_{\rm el}}{\partial \varepsilon_{\Gamma}\partial\varepsilon_{\Gamma'}}.
	\label{eq:moduli}
\end{equation} 
Typically, elastic moduli are written in Voigt notation, which can obscure the connection to the underlying symmetry of the lattice. I will clarify as needed.

\subsection{Strain Coupling in Single-Component Superconductors}
\label{sse:single}

I first describe how strain couples to a single-component superconducting order parameter---an order parameter whose irreducible representation is one dimensional, or, equivalently, is described by a single complex number.\footnote{Here we are referring to the orbital part of the wavefunction, and are ignoring spin-orbit coupling. Spin orbit coupling can be included in a straightforward way and is discussed in e.g. Theuss \textit{et al.} \cite{theussSinglecomponentSuperconductivityUTe22024a}.} Note that this includes many $s$-, $p$-, $d$-, etc-wave superconductors: the point is not whether the order parameter is conventional, but whether it is single or multi component.

The mean-field free energy density for a superconducting order parameter $\psi=\left|\psi\right| e^{i\phi}$ is
\begin{equation}
		f_{\rm sc} = \frac{1}{2}a(T-\Tc)\left|\psi\right|^2 + \frac{1}{4} \beta \left|\psi\right|^4.
	\label{eq:fsc}
\end{equation}
The lowest-order coupling to strain is always quadratic in order parameter because of the complex phase (equivalently, because the free energy must be gauge invariant). Which power of strain appears in the coupling follows from the requirement that the free energy is a scalar (the identity irrep., e.g. $A_{1g}$ in $D_{4h}$. See \autoref{tab:D4h-even-products} for an example product table used to construct allowed terms in the free energy). Strains that change the volume of the lattice but not its symmetry are scalars and thus couple linearly. Shear strains, however, break the lattice symmetry and couple quadratically---the square of the shear strain, $\varepsilon_{\mathcal{s}}^2$, is a scalar (diagonal elements in \autoref{tab:D4h-even-products}). This yields
\begin{equation}
	f_{\rm coup} = \sum_{\Gamma_\mathcal{c}} \gamma_{\Gamma_\mathcal{c}} \varepsilon_{\Gamma_\mathcal{c}}\left|\psi\right|^2 + \sum_{\Gamma_\mathcal{s}} \gamma_{\Gamma_{\mathcal{s}}} \varepsilon_{\Gamma_\mathcal{s}}^2\left|\psi\right|^2,
	\label{eq:coup}
\end{equation}
where $\Gamma_\mathcal{c}$ label scalar strains and $\Gamma_\mathcal{s}$ label shear strains.

\newcommand{\irrep}[1]{\boldmath$\mathbf{#1}$}

\begin{table}[t]
	\centering
	\caption{\textbf{Direct products of the $D_{4h}$ point group.} Constructing allowed terms in the free energy ($A_{1g}$ irrep.) amounts to building products that transform as $A_{1g}$ from objects that transform in non-trivial ways, like strains and order parameters. $R_z$ denotes an axial vector along the $z$ axis, such as torque or magnetic field. For brevity, only even ($g$) irreps. are shown: extending the table to include odd ($u$) irreps. follows the rules $\Gamma_g \otimes \Gamma_u
		=
		\bigl(\Gamma_g \otimes \Gamma_g\bigr)_u$ and $
		\Gamma_u \otimes \Gamma_u
		=
		\bigl(\Gamma_g \otimes \Gamma_g\bigr)_g$, e.g. $B_{2u} \otimes B_{1g}=A_{2u}$.  }
	\label{tab:D4h-even-products}
	
	\renewcommand{\arraystretch}{1.2}
	\setlength{\tabcolsep}{6pt}
	
	\begin{tabular}{c|ccccc|c}
		\toprule
		$\otimes$
		& \irrep{A_{1g}}
		& \irrep{A_{2g}}
		& \irrep{B_{1g}}
		& \irrep{B_{2g}}
		& \irrep{E_g}
		& \textbf{Basis functions}
		\\
		\midrule
		
		\irrep{A_{1g}}
		& $A_{1g}$
		& $A_{2g}$
		& $B_{1g}$
		& $B_{2g}$
		& $E_g$
		& $x^2+y^2,\;z^2$
		\\
		
		\irrep{A_{2g}}
		& $A_{2g}$
		& $A_{1g}$
		& $B_{2g}$
		& $B_{1g}$
		& $E_g$
		& $R_z$
		\\
		
		\irrep{B_{1g}}
		& $B_{1g}$
		& $B_{2g}$
		& $A_{1g}$
		& $A_{2g}$
		& $E_g$
		& $x^2-y^2$
		\\
		
		\irrep{B_{2g}}
		& $B_{2g}$
		& $B_{1g}$
		& $A_{2g}$
		& $A_{1g}$
		& $E_g$
		& $xy$
		\\
		
		\irrep{E_g}
		& $E_g$
		& $E_g$
		& $E_g$
		& $E_g$
		& $A_{1g}\oplus \left[A_{2g}\right]\oplus B_{1g}\oplus B_{2g}$
		& $\left\{xz,\;yz\right\}$
		\\
		
		\bottomrule
	\end{tabular}
\end{table}

With these definitions, we write the total free energy density as
\begin{equation}
	f_{\rm tot} = f_{\rm el}+f_{\rm sc}+f_{\rm coup}.
	\label{eq:tot}
\end{equation}

For clarity, I will treat the cases of shear and scalar strains separately. Considering a single scalar strain, we minimize \autoref{eq:tot} with respect to the order parameter magnitude $\left|\psi\right|$ and find two solutions (the third is related to the second by a phase factor and thus redundant):
\begin{eqnarray}
	\left|\psi\right|_+ =& 0 & \hspace{1em} T>\Tc, 	\label{eq:op1}\\
	\left|\psi\right|_- =& \sqrt{\frac{a(\Tc-T)-\gamma_{\Gamma_\mathcal{c}}\varepsilon_{\Gamma_\mathcal{c}}}{\beta}} & \hspace{1em} T<\Tc.
	\label{eq:op2}
\end{eqnarray}

These solutions can now be substituted into \autoref{eq:tot} to solve for the elastic moduli above and below \Tc:
\begin{eqnarray}
	c_{\Gamma_\mathcal{c},+} = &c_{\Gamma_\mathcal{c}},\\
	c_{\Gamma_\mathcal{c},-} = &c_{\Gamma_\mathcal{c}} - \frac{\gamma_\mathcal{c}^2}{2 \beta}.
	\label{eq:moddif}
\end{eqnarray}
This is typically expressed as the discontinuous change in a scalar modulus across \Tc:
\begin{equation}
	\Delta c_{\Gamma_\mathcal{c}} = -\frac{\gamma_{\Gamma_\mathcal{c}}^2}{2 \beta}.
	\label{eq:jump}
\end{equation}
\autoref{eq:jump} is what we refer to as the ``jump'' in a scalar modulus at \Tc. A similar exercise gives the specific heat jump:
\begin{equation}
	 \frac{\Delta C}{T} =  \frac{a^2}{2 \beta}.
	\label{eq:spec}
\end{equation}
Finally, it is clear from \autoref{eq:op2} that the slope of \Tc with scalar strain is given by 
\begin{equation}
	\frac{d\Tc}{d \varepsilon_{\Gamma_\mathcal{c}}} = -\frac{\gamma_{\Gamma_\mathcal{c}}}{a}.
	\label{eq:slope}
\end{equation}
Combining \autoref{eq:jump}, \autoref{eq:spec}, and \autoref{eq:slope} gives a relation between the jump in specific heat and the jump in elastic modulus:
\begin{equation}
	\Delta c_{\Gamma_\mathcal{c}} = -\left(	\frac{d\Tc}{d \varepsilon_{\Gamma_\mathcal{c}}}\right)^2  \frac{\Delta C}{T}.
	\label{eq:ehrenfest}
\end{equation}
In words, the jump in elastic modulus is equal---but of opposite sign---to the jump in specific heat times the square of the derivative of \Tc with strain. This is a powerful consistency check between different experiments on the same material. It also allows one to use two known quantities to predict the third if it has not yet been measured.

\autoref{eq:ehrenfest} is more general than what we have derived here using Landau mean field theory: \autoref{eq:ehrenfest} is an Ehrenfest relation that only requires that the first derivatives of the free energy are continuous across \Tc, i.e. that the transition is second order, which is almost always the case for superconductors in zero magnetic field.

For shear strains, the algebra is similar except that the order parameter below \Tc is given by
\begin{equation}
\left|\psi\right|_- =	\sqrt{\frac{a(\Tc-T)-\gamma_{\Gamma_\mathcal{s}}\varepsilon_{\Gamma_\mathcal{s}}^2}{\beta}}.
	\label{eq:ops}
\end{equation}
Solving for the change in elastic modulus, we find
\begin{equation}
	\Delta c_{\Gamma_\mathcal{s}} = -\frac{a(T-\Tc)\gamma_{\Gamma_\mathcal{s}}}{\beta}.
	\label{eq:slope2}
\end{equation}
Now there is no jump immediately at \Tc, but instead a change in slope whose sign depends on the sign of the coupling, $\gamma_{\Gamma_\mathcal{s}}$, between shear strain and superconductivity.

That's it for single-component superconductors: scalar moduli have jumps at \Tc, whereas shear moduli have ``kinks''. While measurements of these features yield valuable information about the coupling between order parameter and strains, no specific insight into the symmetry of the superconducting order parameter itself is gained: $s$, $p$, $d$, spin singlet, and spin triplet order parameters---indeed, all second-order phase transitions---exhibit the same qualitative phenomenology \footnote{One exception is when the order parameter has all the same symmetry properties (i.e. is of the same irrep.) as one of the strains. Then, there is bilinear coupling between strain and order parameter and the associated elastic modulus exhibits divergent, Curie-Weiss-like behavior \cite{rehwaldUltrasonicPropertiesStrontium1971,ramshawAvoidedValenceTransition2015}.}.

This will change for multi-component superconductors.  

\subsection{Strain Coupling in Multi-Component Superconductors}
\label{sse:multi}


The mean-field free energy density of a symmetry-enforced, two-component superconducting order parameter in a system with $C_4$ rotation symmetry (e.g. $D_{4h}$) is \cite{sigristEhrenfestRelationsUltrasound}
\begin{equation}
\begin{aligned}
	f_{\rm sc} = \frac{1}{2}a(T-\Tc)\left(\left|\psi_1\right|^2+\left|\psi_2\right|^2\right) &+ \frac{1}{4} \beta_1 \left(\left|\psi_1\right|^2+\left|\psi_2\right|^2\right)^2 + \frac{1}{4}\beta_2\left(\left|\psi_1\right|^2-\left|\psi_2\right|^2\right)^2\\
&+\frac{1}{4}\beta_3\left({\psi_1^{\star}}\psi_2+{\psi_2^{\star}}\psi_1\right)^2,
	\label{eq:fscmulti}
\end{aligned}
\end{equation}
where the two-component order parameter is $\bm{\psi}=\left\{\left|\psi_1\right| e^{i \phi_1},\left|\psi_2\right| e^{i \phi_2}\right\}$. The term with the coefficient $\beta_1$ depends only on the magnitude of the order parameter. The $\beta_2$ term is an anisotropy term, and the $\beta_3$ term depends on the relative phase of the two components. Stability of the system requires that $\beta_1>0$, $\beta_1+\beta_2>0$, and $\beta_1+\beta_3>0$ \footnote{\autoref{eq:fscmulti} is written for a tetragonal point group. There are more independent quartic coefficients in low-symmetry point groups, and fewer in higher symmetry point groups (i.e. $\beta_2$ and $\beta_3$ are related by symmetry in hexagonal point group)}.

\autoref{eq:fscmulti} can be minimized with respect to $\psi_1$, $\psi_2$, and the phase difference $\phi \equiv \phi_1 - \phi_2$ to find the ground state. Taking the concrete example of the $p$-wave state in a tetragonal crystal, when $\beta_2<0$ and $\beta_2<\beta_3$, the ground state is the nematic $\psi_x$ or $\psi_y$ (any overall sign can be absorbed into the absolute phase of the order parameter). When $\beta_3<0$ and $\beta_3<\beta_2$, then the ground state is the diagonal nematic $\psi_x\pm \psi_y$. Finally, if $\beta_2>0$ and $\beta_3>0$ then the ground state has a phase difference $\phi = \pi/2$ between the two components and forms the chiral state $\psi_x \pm i \psi_y$.

The couplings to strains also change for a two-component order parameter. The general motif is that gauge-invariant bilinears of the superconducting order parameter can now be formed that couple to shear strain. This leads to jumps in shear elastic moduli at \Tc, which are forbidden for single-component order parameters. Which couplings are allowed depends on the symmetry of the lattice and the particular multi-component order parameter in question, and therefore a ``general" Landau theory is both cumbersome and uninstructive. 

Instead, I will give a worked example of the strain couplings for a two-component order parameter that is particularly relevant for many contemporary superconductors. More examples will be discussed as needed in \autoref{se:experiments}.

\subsubsection{$p_x+ip_y$ in a tetragonal crystal}
\label{sse:tetrag}

The obvious case to discuss is the $p_x +i p_y$ state in a tetragonal crystal structure. To be consistent with our earlier notation, we will refer to the order parameter components as $\psi_x$ and $\psi_y$. In a tetragonal crystal, the $l = 1$ angular momentum channel splits into one singlet---$\psi_z$ of the $A_{2u}$ irrep.---and one doublet---$\bm{\psi}_{\rm E_u}=\left\{\psi_x,\psi_y\right\}$ of the $E_u$ irrep. It is the $E_u$ doublet that I consider here. 

To construct the free energy density that describes the coupling to strain, we first want to identify the gauge-invariant (i.e. real) bilinears of $\bm{\psi}_{\rm E_u}$. There are three: $\left|\psi_x\right|^2+\left|\psi_y\right|^2$, $\left|\psi_x\right|^2-\left|\psi_y\right|^2$, and ${\psi_x^{\star}}\psi_y+{\psi_y^{\star}}\psi_x=2\left|\psi_x\right|\left|\psi_y\right|\cos\left(\phi_x-\phi_y\right)$, of the $A_{1g}$, $B_{1g}$, and $B_{2g}$ irreps., respectively. Physically, these three bilinears correspond to the overall amplitude of the order parameter, the relative amplitudes of the two components, and the relative phase of the two components. The latter two are new degrees of freedom---not present for single-component order parameters---that can couple to shear strain. 

The couplings are now straightforward to identify: the $A_{1g}$ bilinear couples to $A_{1g}$ strains, the $B_{1g}$ bilinear couples to $B_{1g}$ strain, etc. We have
\begin{equation}
\begin{aligned}
	f_{\rm coup} =& \gamma_{A_{1g,1}} \left(\varepsilon_{xx}+ \varepsilon_{yy}\right)\left(\left|\psi_x\right|^2+\left|\psi_y\right|^2\right)+\gamma_{A_{1g,2}} \varepsilon_{zz}\left(\left|\psi_x\right|^2+\left|\psi_y\right|^2\right)\\
	            & \gamma_{B_{1g}} \left(\varepsilon_{xx}- \varepsilon_{yy}\right)\left(\left|\psi_x\right|^2-\left|\psi_y\right|^2\right)+2\gamma_{B_{2g}} \varepsilon_{xy} \left|\psi_x\right|\left|\psi_y\right|\cos\left(\phi_x-\phi_y\right).
	\label{eq:multifen}
\end{aligned}
\end{equation}
The mechanics of solving for the jumps are relatively straightforward but cumbersome (see Ghosh \textit{et al.} \cite{ghoshThermodynamicEvidenceTwocomponent2021} and Theuss \textit{et al.} \cite{theussSinglecomponentSuperconductivityUTe22024a} for details). 

The first two terms in \autoref{eq:multifen} give jumps in the scalar elastic moduli $(c_{11}+c_{12})/2$ and $c_{33}$, respectively, as well as a jump in the cross-coupling modulus $c_{13}$. As I will discuss in the context of \sro, the fact that there are two coupling constants governing three jumps means that the jumps are not independent of one another and can be used with Ehrenfest relations as a consistency check (this fact is true for both single and multi-component order parameters).

The second two terms in \autoref{eq:multifen} give jumps in the shear elastic moduli $(c_{11}-c_{12})/2$ and $c_{66}$, respectively. This result depends only on the symmetry of the lattice and the symmetry of the order parameter in question---it is independent of the microscopic mechanism of pairing, and even independent of \textit{which} particular ordered state forms (i.e. the nematic and chiral ordered states all have jumps in shear moduli at \Tc). The actual \textit{magnitudes} of the jumps, however, depend on the particular ordered state: they are of the form $\Delta c/c \propto \gamma^2/\beta$, where the particular $\gamma$ and $\beta$ depend on the particular ordered state and elastic modulus being measured (again, see Ghosh \textit{et al.} \cite{ghoshThermodynamicEvidenceTwocomponent2021} for details).

\section{EXPERIMENTS}
\label{se:experiments}

The symmetry arguments of the previous section are relatively straightforward---does this program work in reality? 

The clearest case of a jump in a shear modulus from a multi-component order parameter is at the second-order structural transition in SrTiO$_3$. There, the order parameter corresponds to the rotation angle of the TiO$_6$ octahedra and forms a three-dimensional representation---$T_{1g}$ irrep. of the $O_h$ point group. Linear coupling of the order parameter to strain is prohibited both because there are no strains in the $T_{1g}$ irrep., but also because the order parameter forms at finite-$Q$ and the strains used in ultrasound are long wavelength ($Q\approx0$). The square of this order parameter, however, couples to all 6 strains: the $A_{1g}$ scalar strain, but also the $E_g$ and $T_{2g}$ shear strains. As predicted, there are clear jumps in the shear moduli $(c_{11}-c_{12})/2$ ($E_g$) and $c_{44}$ ($T_{2g}$), as well as in the scalar modulus \cite{rehwaldUltrasonicPropertiesStrontium1971,rehwaldAnomalousUltrasonicAttenuation1970}. Other examples include Fe$_{0.92}$O and MnO, where the antiferromagnetic order parameter couples to rhombohedral shear strain (i.e. $\epsilon_{xy}$, $\epsilon_{xz}$, and $\epsilon_{yz}$), producing a jump in $c_{44}$ at the Neel transition \cite{suminoELASTICCONSTANTSSINGLE1980}.

These physical examples demonstrate that the coupling constants to the shear strains are of order the same size as the coupling constants to scalar strains, at least for structural and magnetic phase transitions. The remaining questions are 1) whether the numbers work out for superconductors and 2) whether there \textit{are} any multi-component superconductors. Here I review the experimental progress so far. 

\subsection{\sro}
For over 20 years, \sro was the best material candidate for realizing spin-triplet, odd-parity, $p_x\pm ip_y$ superconductivity. The strongest evidence for this state came from NMR, which showed a constant spin susceptibility across \Tc that is indicative of spin-triplet pairing \cite{ishidaSpintripletSuperconductivitySr2RuO41998}. Coupled with various reports of time-reversal symmetry breaking \cite{lukeTimereversalSymmetrybreakingSuperconductivity1998,xiaHighResolutionPolar2006}, the case for $p_x+ip_y$ was compelling. 

This changed in 2019, when Pustogow \textit{et al.} \cite{pustogowConstraintsSuperconductingOrder2019} showed that \sro exhibited a drop in the spin susceptibility consistent with singlet pairing. Subsequent work \cite{chronisterEvidenceEvenParity2021,ishidaReduction17OKnight2020} made the case essentially air-tight: \sro is a spin-singlet superconductor, and therefore odd-parity $p$-wave is almost certainly out of the question. 

This left open the question of where the apparent time-reversal symmetry breaking was coming from---naively, the only way to break time reversal in the superconducting state is to form a chiral superposition of two components. If $p_x \pm i p_y$ is ruled out, then what else could break TRS?

This set the stage for a pair of papers published in 2021 that reported jumps in the $c_{66}$ shear elastic modulus at \Tc in \sro \cite{ghoshThermodynamicEvidenceTwocomponent2021,benhabibUltrasoundEvidenceTwocomponent2021}, suggesting a multi-component order parameter (these results were preceded by reports of a jump in $c_{66}$ by Okuda \textit{et al.} \cite{okudaUnconventionalStrainDependence2002}, although interpretation of those results was complicated by a jump in $c_{44}$, which is forbidden by symmetry for any order parameter in $D_{\rm 4h}$).

\begin{figure}[ht!]
	\centering
	\includegraphics[width=0.95\textwidth]{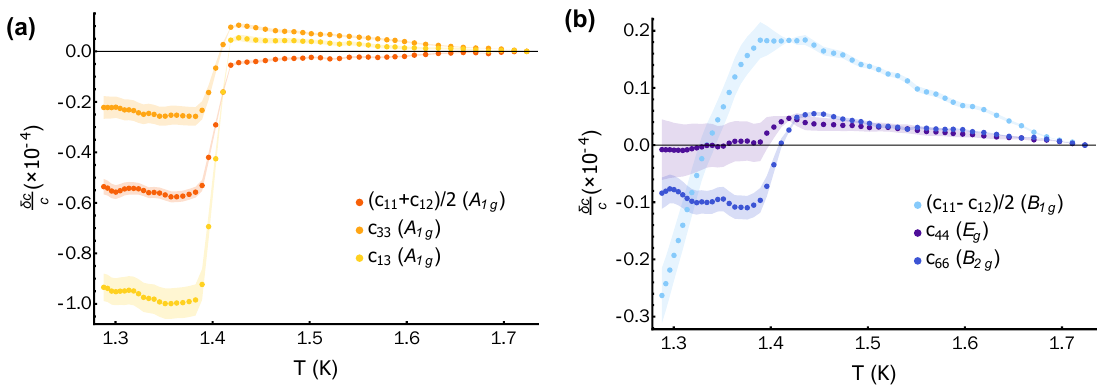}
	\caption{\textbf{Relative change in elastic moduli through \Tc of \sro.} Scalar (a) and shear (b) moduli of \sro across \Tc, along with the experimental errors that arise from uncertainties in sample dimensions, represented by shaded regions. All data were taken simultaneously (i.e. all six moduli were acquired at the same time for each temperature point) using resonant ultrasound spectroscopy. The jump in the $c_{66}$ shear modulus is indicative of a two-component order parameter. From \cite{ghoshThermodynamicEvidenceTwocomponent2021}.  }
	\label{fig:sr2ruo4}
\end{figure}

The study of Ghosh \textit{et al.} \cite{ghoshThermodynamicEvidenceTwocomponent2021} used resonant ultrasound spectroscopy (RUS): a technique that determines all elastic moduli at once by fitting a broad mechanical resonance spectrum. They reported a jump in the $c_{66}$ shear modulus of $\Delta c_{66}/c_{66} \approx 2\times10^{-5}$ (see \autoref{fig:sr2ruo4}). They also measured jumps in the scalar moduli $(c_{11}+c_{12})/2$, $c_{33}$, and $c_{13}$. As noted in \autoref{sse:tetrag}, these three jumps should be related to one another because they arise from only two coupling coefficients. Specifically, the relation is
\begin{equation}
\Delta c_{13}^2 = \Delta  (c_{11}+c_{12})/2 \times \Delta c_{33}.
\end{equation}
Empirically, Ghosh \textit{et al.} found that $\Delta (c_{11}+c_{12})/2 \times \Delta c_{33} = (9.9 \pm 1.5)\times 10^{-5}$ GPa$^2$, and $\Delta c_{13}^2 = (8.3 \pm 1.1)\times 10^{-5}$ GPa$^2$. This consistency check provides some confidence that the measured jumps are not an experimental artifact (disagreement between these jumps and the measured specific heat jump using Ehrenfest relations is still unresolved, however \cite{ghoshThermodynamicEvidenceTwocomponent2021}).

Concurrently, Benhabib \textit{et al.} \cite{benhabibUltrasoundEvidenceTwocomponent2021} used pulse-echo ultrasound to measure a jump in $c_{66}$ of $\Delta v_{66}/v_{66} \approx 2\times10^{-7}$, or $\Delta c_{66}/c_{66} \approx 4\times10^{-7}$---50 times smaller than what was measured by Ghosh \textit{et al.} Measurements of the other moduli were not reported. 

At the time, the quantitative disagreement between the two experiments was attributed to the two measurements being performed at very different frequencies---2 MHz for RUS, versus 169 MHz for pulse echo. This would require some relatively slow relaxational dynamics in the system, with the relaxation rate intermediate to the two measurement frequencies. While such dynamics have been hinted at in a separate RUS study of \sro \cite{ghoshStrongIncreaseUltrasound2022}, this is in some sense a fine-tuned argument. More recently, unpublished work using the same sample as the RUS study, but employing pulse-echo ultrasound, has confirmed that the jumps in $c_{33}$ and $c_{66}$ are independent of frequency up to 3 GHz, and that the magnitude of these jumps is the same as was found with RUS \cite{shragai2026}. These measurements are therefore all in the thermodynamic limit, and thus there is no justification for invoking frequency-dependent effects to explain discrepancies between Ghosh \textit{et al.} and Benhabib \textit{et al.}. This suggests that there may be sample variation between the two studies, or that there is some experimental artifact in one of the measurements that is unaccounted for.

Putting aside the quantitative difference, both measurements agree that \sro appears to host a multi-component order parameter. Immediate options are order parameters in the $E_u$ irrep.---which notably contains the $p_x+ip_y$ state---and the $E_g$ irrep., which contains the $d_{xz}+id_{yz}$ state. With NMR ruling out spin triplet and thus odd parity, it would seem that $d_{xz}+id_{yz}$ is the state most consistent between the NMR and ultrasound experiments. Further, such a state would explain the broken time reversal observed by $\mu$SR and polar Kerr rotation \cite{lukeTimereversalSymmetrybreakingSuperconductivity1998,xiaHighResolutionPolar2006}.

There are several issues with this interpretation that are currently unresolved:
\begin{enumerate}
	\item A jump in $c_{66}$ implies coupling of the form $\varepsilon_{xy} \left|\psi_{xz}\right|\left|\psi_{yz}\right|\cos\left(\phi_{xz}-\phi_{yz}\right)$, which immediately implies that \Tc should shift \textit{linearly} with applied shear strain, i.e. $\Tc \propto \left|\varepsilon_{xy}\right|$. Moreover, the single \Tc should split into two \Tc's with $\varepsilon_{xy}$. Such linear coupling and splitting has been looked for but not observed \cite{jerzembeck$T_c$ElastocaloricEffect2024,mattoniDirectEvidenceAbsence2025}.
	
	\item There is no jump in $(c_{11}-c_{12})/2$, which is also allowed by symmetry for the $d_{xz}+id_{yz}$ state (although there is no constraint on the size of the effect). There is also no corresponding linear coupling to \Tc in the $\varepsilon_{xx}-\varepsilon_{yy}$ channel \cite{liHighsensitivityHeatcapacityMeasurements2021} (this result is consistent with no jump in $(c_{11}-c_{12})/2$, although it seems to rule against a $d_{xz}+id_{yz}$ state.)
	
	\item The $d_{xz}+id_{yz}$ state has a horizontal line node at $k_z = 0$---no pairing at zero momentum along the $c$ axis. This implies some sort of inter-layer pairing mechanism, which would be unusual given the highly 2D nature of the electronic structure in \sro. It would also be inconsistent with claims of vertical line nodes from thermal transport \cite{hassingerVerticalLineNodes2017}, critical field measurements \cite{landaetaEvidenceVerticalLine2024}, ultrasonic attenuation \cite{lupienUltrasoundAttenuation$mathrmSr_2mathrmRuO_4$2001}, and quasiparticle interference \cite{sharmaMomentumresolvedSuperconductingEnergy2020}.
\end{enumerate}

This is where the current understanding of \sro stands: two independent ultrasound measurements suggest a two-component superconducting order parameter, but a number of follow up experiments have failed to find the implied shear strain coupling. One proposed solution invokes an accidentally-degenerate $d_{x^2-y^2}+ i g_{xy(x^2-y^2)}$ order parameter \cite{kivelsonProposalReconcilingDiverse2020}. This state solves the second two issues: it \textit{only} has a jump in $c_{66}$, and it has vertical line nodes that mimic those of $d_{x^2-y^2}$. However, it fails to deal with the first issue---the absence of linear coupling of \Tc to shear strain---and it introduces a new issue: the issue of fine tuning (it naturally has two \Tc's for the two separate components).

At present, it is my opinion that the solution might be rather unsavory: meso-scale inhomogeneity within the samples (for example, the inclusion of Sr$_3$Ru$_2$O$_7$ layers) might be smearing out the implied linear coupling, breaking symmetry in a way such that the jump in $c_{66}$ is actually coming from a scalar modulus jump, or introducing a secondary order parameter (such as magnetism) that, for some reason, onsets at \Tc. Current ongoing work is looking at the effect of deliberately introducing Sr$_3$Ru$_2$O$_7$ inclusions on the jump in $c_{66}$.

\subsection{\ute}
\label{sse:ute}
The story of multi-component superconductivity in \ute has two parts: the original claim \cite{hayesMulticomponentSuperconductingOrder2021}, which is now largely understood to be due to sample inhomogeneity, and the second claim \cite{kamatThermodynamicDiscoveryTetracriticality2026}, which is still actively being explored. Crucial for understanding this story is the fact that \ute is orthorhombic ($D_{2h}$ point group); there are only single-component representations available in this group, and thus any two-component order parameter should have two \Tc's. 

The original claim for multi-component superconductivity in \ute came from the observation of the polar Kerr effect---indicative of TRS breaking---below \Tc \cite{hayesMulticomponentSuperconductingOrder2021}. Importantly, this observation was accompanied by the report of two \Tc's split by approximately 80 mK---generally expected for multi-component superconductivity in an orthorhombic setting, where there are no multi-dimensional irreps. The expectation would be that a single order parameter, $\Psi_1$, onsets at the upper transition, and a second order parameter, $\Psi_2$ appears at the lower transition with a $\pi/2$ phase shift, forming the $\Psi_1\pm i \Psi_2$ state. Whether TRS was broken at the upper or lower transition was not resolvable in the experiment, however.

The primary issue with this claim of two-component superconductivity was that some samples---particularly those with higher transition temperatures---only showed a single \Tc \cite{rosaSingleThermodynamicTransition2022}. This led to the suggestion that it was perhaps sample inhomogeneity that was causing the two transitions in some samples. Intriguingly, however, there were typically \textit{only} ever two transitions---not three, four, or more, and not a broadly smeared \Tc. This left open the possibility that something more complex was going on. 

The fact that there are only single-component irreps. in an orthorhombic crystal turns out to be something of an advantage for ultrasound: any pair of the four proposed odd-parity irreps for the superconducting state---$A_u$, $B_{1u}$, $B_{2u}$, and $B_{3u}$---couple to one (and only one) of the shear strains---$B_{1g}$, $B_{2g}$, and $B_{3g}$. This implies that \textit{any} odd-parity, two component state should have a jump in a shear modulus, and which particular shear modulus narrows the options down to 2 pairs. 

\begin{figure}[ht!]
	\centering
	\includegraphics[width=0.95\textwidth]{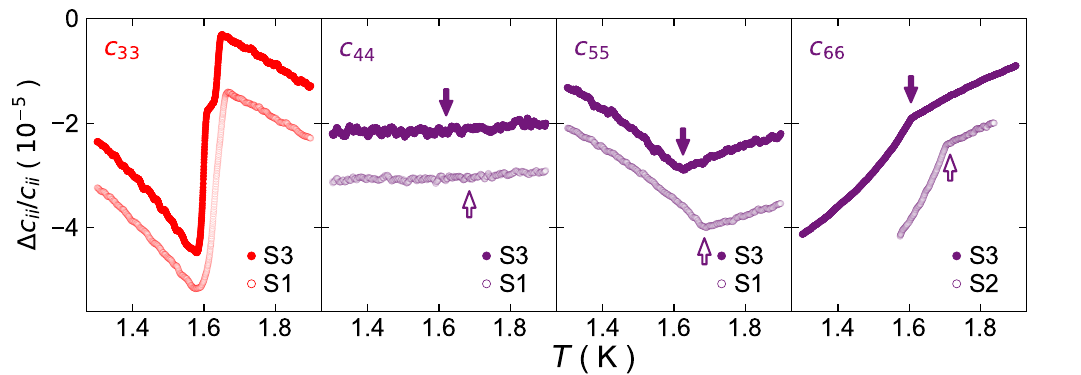}
	\caption{\textbf{Shear moduli in single and double transition \ute samples} The scalar elastic modulus $c_{33}$ shows two distinct discontinuities at \Tc, consistent with the two peaks found in the specific heat measured in the same sample. The shear moduli, on the other hand, show no discontinuities and behave nearly identically to the shear moduli of the single-\Tc sample. Single-(double-)transition samples are shown with empty (filled) symbols. Empty (filled) arrows mark the superconducting transition for all shear moduli for single-(double-)transition samples. Curves have been offset vertically for clarity. From \cite{theussSinglecomponentSuperconductivityUTe22024a}.  }
	\label{fig:UTe2_no_jumps}
\end{figure}

This was investigated by Theuss \textit{et al.} \cite{theussSinglecomponentSuperconductivityUTe22024a}, who used pulse-echo ultrasound to measure 3 scalar and 3 shear moduli in samples that exhibited both single and double superconducting transitions. As shown in \autoref{fig:UTe2_no_jumps}, neither type of sample exhibits a shear-modulus jump at \Tc. This is in spite of the very large coupling to scalar strain, as exhibited by the large jump in $c_{33}$ (and large jumps at both \Tc's in the double-\Tc sample). This rules out \textit{all} multi-component order parameters where the two components are both odd (or both even) parity. Mixed-parity order parameters such as $B_{1g}+ i B_{3u}$---which are not expected for \ute but could exist in principle---are not ruled out by this experiment because the product of an even and an odd irrep. is an odd irrep., but all strains are even (see caption to \autoref{tab:D4h-even-products}).

While \ute appears to be single-component at ambient pressure (a phase I will refer to as SC1), there is clear evidence for a second superconducting phase under pressure (SC2), and two additional superconducting phases in an applied magnetic field \cite{ranExtremeMagneticFieldboosted2019}. It is generally accepted that the pressure-induced SC2 phase is connected to one of the field-induced phases in the $B-P-T$ phase diagram \cite{vasinaConnectingHighFieldHighPressure2025}. The SC3 phase, which occurs only above 35 tesla, is disconnected from the two lower-field phases. 

\begin{figure}[ht!]
	\centering
	\includegraphics[width=0.95\textwidth]{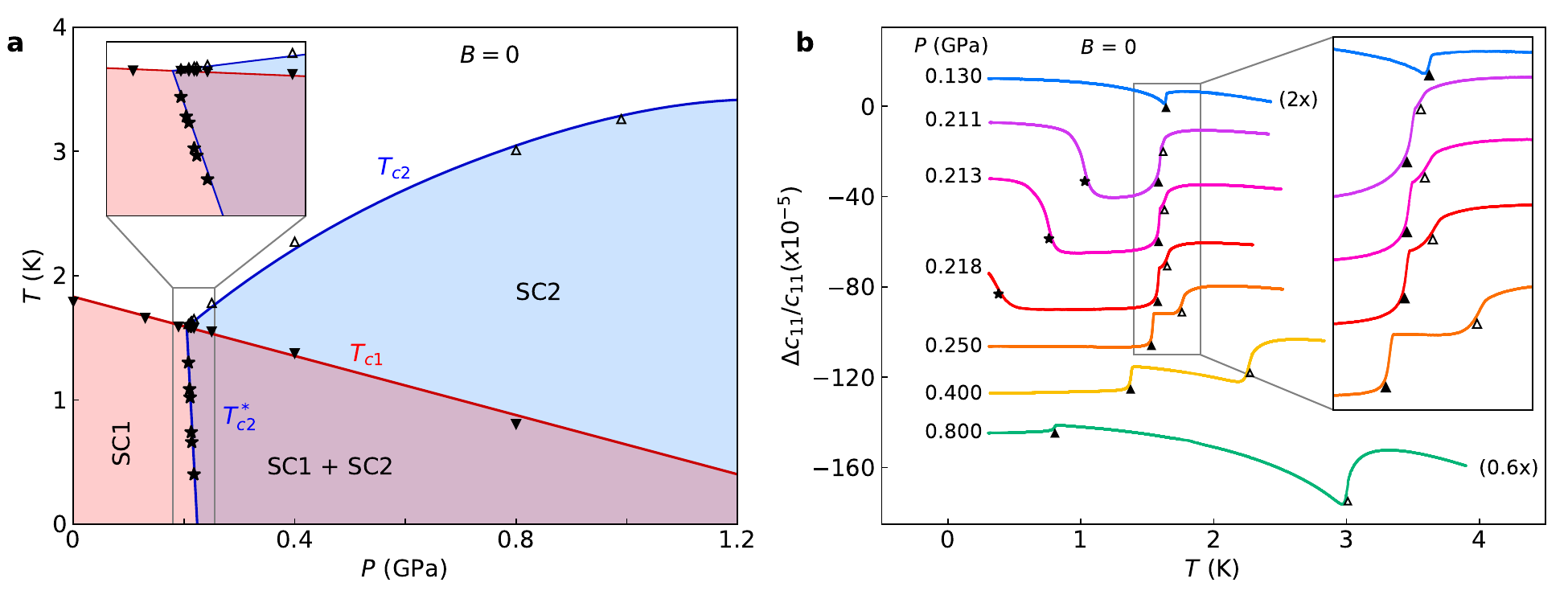}
	\caption{\textbf{Phase diagram of \ute under pressure} \textbf{a}) $T_{\rm c1}$, $T_{\rm c2}$ and $T_{\rm c2}^{\star}$ are superconducting transitions represented by solid triangles, hollow triangles, and a star respectively. The solid red and blue lines are guides to the eye that represent the SC1 and SC2 phase boundaries, respectively. The purple-shaded SC1+SC2 region is a multi-component state. b) The  $c_{33}$ elastic modulus. For $P<0.2$ GPa, there is a downward jump at $T_{\rm c1}$ when \ute enters the SC1 state. This jump grows for $P>0.2$ GPa, and occurs at $T_{\rm c2}$. At $P=0.21$ GPa, there is an upward jump at $T_{\rm c2}^{\star}$ as the sample exits the SC2 superconducting state upon cooling. From \cite{kamatThermodynamicDiscoveryTetracriticality2026}.  }
	\label{fig:UTe2_phaseD}
\end{figure}

Do SC1 and SC2 interact in some region of parameter space to form a multi-component state? Until recently, the proposed pressure-temperature phase diagram of \ute contained a thermodynamically-forbidden tricritical point where the SC2 phase intersects the SC1 phase near $P = 0.20$ GPa (\autoref{fig:UTe2_phaseD}) \cite{vasinaConnectingHighFieldHighPressure2025}. Generally, such an intersection must be a \textit{tetra}critical point if all phase boundaries are 2nd order. Various solutions were proposed, including a hidden, fourth phase boundary, or a 3rd-order transition line for the SC2 phase.

Kamat \textit{et al.} \cite{kamatThermodynamicDiscoveryTetracriticality2026} recently resolved this puzzle by discovering a fourth transition line that bends backwards at the intersection of SC1 and SC2: SC2 is perturbed so strongly by the presence of SC1 that it disappears upon \textit{cooling}. One striking feature of the data is a jump \textit{up} at the transition where SC2 is lost on cooling ($T_{c2}^{\star}$ in \autoref{fig:UTe2_phaseD}). As discussed in \autoref{sse:single}, the jump in elastic modulus is required to have the opposite sign of the specific heat jump (\autoref{eq:ehrenfest}). This implies a \textit{negative} specific heat jump---something previously never reported for a superconducting transition. However, this is consistent with the structure of the phase boundary: upon cooling through $T_{c2}^{\star}$, the system \textit{loses} the SC2 component of the order parameter, resulting in a rate of entropy increase that is higher in the low-temperature SC1 phase than in the intermediate temperature SC1+SC2 phase. 

The revised $P-T$ (and $B_b-P-T$ \cite{kamatThermodynamicDiscoveryTetracriticality2026}) phase diagram identifies a region of parameter space where two distinct order parameters are not only present, but also are strongly interacting with one another. What are the two order parameters, and do they form a chiral, TRS-breaking state? This is still an unresolved question. To date, measurements of the $c_{55}$ shear modulus ($B_{2g}$ irrep.) across the SC2 to SC1+SC2 phase boundary have been reported, and show no jump. Taken at face value, this rules out the $B_{1u}+iB_{3u}$ and $A_u + i B_{2u}$ combinations (it also rules out non-TRS-breaking combinations of those pairs). Future measurements of $c_{66}$ ($B_{1g}$) will inform on the $B_{2u}+iB_{3u}$ and $A_u + i B_{1u}$ combinations, while measurements of $c_{44}$ ($B_{3g}$) will inform on the $B_{1u}+iB_{2u}$ and $A_u + i B_{3u}$ combinations. Direct measurements of TRS breaking---such as $\mu$SR or the field-dependence of the elastic moduli---should also be pursued. As noted in Kamat \textit{et al.} \cite{kamatThermodynamicDiscoveryTetracriticality2026}, the SC1+SC2 phase also exists in an applied magnetic field at ambient pressure, but there the determination of TRS breaking is complicated by the explicit TRS breaking of the external magnetic field.

\subsection{\upt}

\upt has long been accepted as the best established two-component superconductor outside of $^3$He (a superfluid, strictly speaking). Two superconducting transitions---at approximately 350 and 450 mK, depending on sample quality---are clearly observed in bulk specific heat measurements, and the onset of TRS breaking is observed at the lower of these two transitions \cite{schemmObservationBrokenTimereversal2014}. Coupled with Knight shift \cite{touOddParitySuperconductivityParallel1996} and Josephson tunneling experiments \cite{strandTransitionRealComplex2010}, the general consensus is that \upt hosts a chiral $(p_x^2-p_y^2)p_z + i (2p_x p_y)p_z$ order parameter ($E_{2u}$ irrep. of $D_{6h}$). 

\begin{figure}[ht!]
	\centering
	\includegraphics[width=0.6\textwidth]{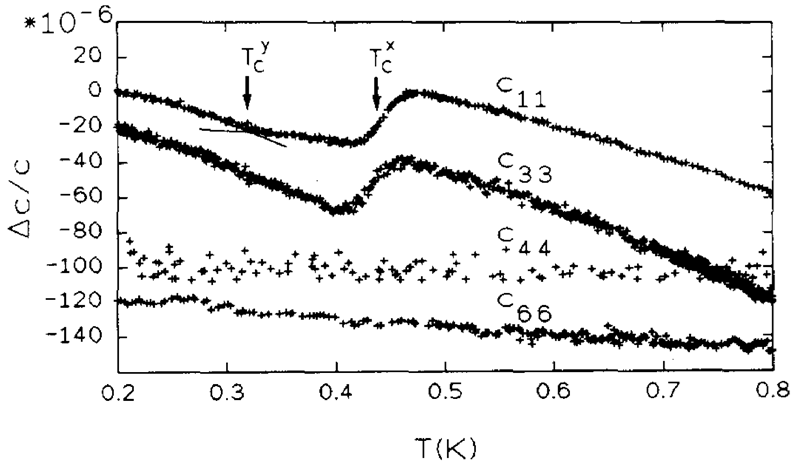}
	\caption{\textbf{Elastic moduli of \upt through \Tc.} $c_{11}$ and $c_{33}$ exhibit sharp drops at the upper transition temperature as expected. There is no clear signature of the lower transition; there is a small kink in $c_{11}$ that could be consistent with a second transition. A jump in $c_{66}$ is allowed by symmetry at the lower \Tc if the order parameter is the proposed $E_{2u}$, $f$-wave state. From \cite{thalmeierElasticConstantAnomalies1991}.    }
	\label{fig:upt3}
\end{figure}

The proposed two-component $E_{2u}$ order parameter can form a bilinear that couples directly to shear strain (both $\varepsilon_{xx}-\varepsilon_{yy}$ and $\varepsilon_{xy}$, of the two-dimensional $E_{2g}$ irrep.) The associated elastic modulus, $c_{66}$, was investigated by Thalmeier \textit{et al.} \cite{thalmeierElasticConstantAnomalies1991}, but no jump at either \Tc was observed (see \autoref{fig:upt3}). The authors attributed this to a small coupling coefficient, and not as evidence against a two-component order parameter. It is worth noting that the resolution of their $c_{66}$ measurement is around 10 ppm---roughly two orders of magnitude lower than state of the art---warranting a re-investigation of this modulus with higher resolution measurements and in newer, higher-quality samples.  

A crucial question is why \upt should have two \Tc's at all: the proposed two order parameter components are degenerate in $D_{6h}$, and should therefore condense at a single \Tc. The splitting of the two \Tc's has sometimes been attributed to the presence of symmetry-breaking antiferromagnetic order that is proposed to onset at $T_N = 5$ K. Coupling of the magnetism to the lattice (which is always present) then lowers the symmetry of the lattice and splits the degenerate $E_{2u}$ irrep. 

The magnetism in \upt is, however, not static \cite{gannonSpinSusceptibilityTopological2017}, and there is no convincing thermodynamic evidence for an antiferromagnetic phase transition. The fact that the two \Tc's depend on how the sample is prepared, and the more recent experience with two transitions in \ute discussed in \autoref{sse:ute}, suggests that sample inhomogenetiy may also be an issue. A more recent set of polarized neutron scattering experiments has concluded that the spin susceptibility may not be a reliable indicator of the spin state of the Cooper pairs \cite{gannonSpinSusceptibilityTopological2017}. This, plus the unresolved origin of the transition splitting, plus the apparent lack of jump in $c_{66}$, leaves the question of the \upt order parameter very much unresolved. 

\subsection{\bkfa}

Superconductivity in optimally-doped iron pnictides is generally understood to be of the $s_{\pm}$ variety \cite{hirschfeldGapSymmetryStructure2011}. That is, there are $s$-wave gaps on both the hole and electron Fermi surfaces that have a $\pi$ phase difference between them. Upon hole doping, the electron pocket eventually disappears, and some experiments suggest that the system then transitions to a $d$-wave state \cite{maitiEvolutionSuperconductingState2011,thomaleExoticDWaveSuperconducting2011,reidDwaveSwavePairing2012}. Although unconventional in the sense that there is a sign-changing order parameter, these order parameters are all single component in the sense of irreps. of the point group of \bkfa ($D_{4h}$).

\begin{figure}[ht!]
	\centering
	\includegraphics[width=0.65\textwidth]{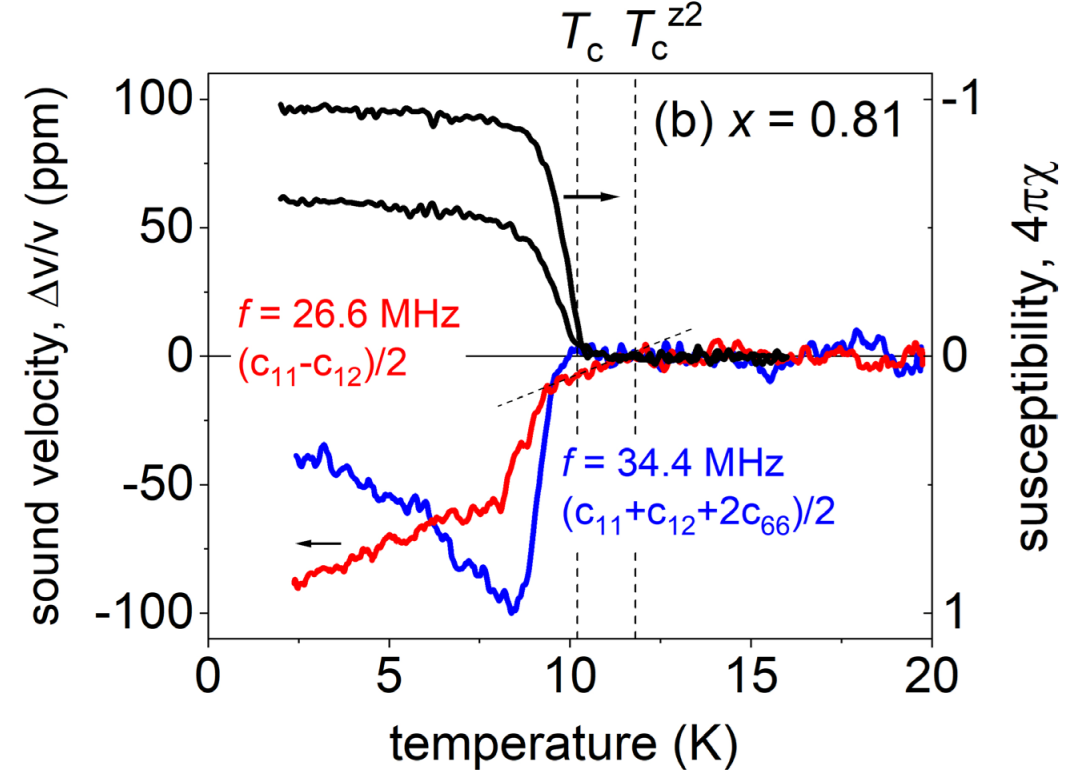}
	\caption{\textbf{Longitudinal and shear elastic moduli of Ba$_{0.19}$K$_{0.81}$Fe$_2$As$_2$.} The magnetic susceptibility shows a clear transition only at \Tc, as does the longitudinal mode $(c_{11}+c_{12}+2c_{66})/2$. The shear mode $(c_{11}-c_{12})/2$ shows a kink at the upper $T_{\rm c}^{\rm Z2}$ and then a jump at \Tc, possibly indicative of a multi-component state. From \cite{grinenkoStateSpontaneouslyBroken2021}.   }
	\label{fig:bafeas}
\end{figure}

There are, however, $\mu$SR experiments that report broken time reversal symmetry in a narrow range of doping---between x = 0.7 and 0.85---in Ba$_{1-x}$K$_x$Fe$_2$As$_2$ \cite{grinenkoStateSpontaneouslyBroken2021}, potentially associated with the transition from $s_{\pm}$ to $d$ (or to a different $s$-wave state). The authors of Grinenko \textit{et al.} \cite{grinenkoStateSpontaneouslyBroken2021} reported a jump in the shear $(c_{11}-c_{12})/2$ modulus, as well as the expected jump in the $(c_{11}+c_{12}+2c_{66})/2$ longitudinal mode. In addition, they report a kink in the shear modulus at an upper transition, dubbed $T_{\rm c}^{\rm Z2}$ in their notation (see \autoref{fig:bafeas}). 

The authors interpret these two features (along with $\mu$SR, specific heat, resistivity, and Nernst effect data) in terms of a two-component superconducting phase, where the two components are either of the same symmetry but on different bands (i.e. $s$ pairing on the electron and hole bands), or are two accidentally-degenerate pairing states of different irreps. (i.e. $s$ and $d$.) They propose that at the upper of the two transitions---$T_{\rm c}^{\rm Z2}$---the relative phase between the two components is locked at $\pm \pi/2$, breaking time reversal symmetry but still having a resistive (though bosonic) state. Then, at the lower \Tc, the global $U(1)$ symmetry is broken and the system develops a Meissner effect and $R = 0$. While this would not be a chiral topological superconductor, it is a related state that shares similar experimental challenges in its detection. 

Follow up studies by the same group at a similar doping found a slightly different set of results: jumps in $c_{11}$ and $c_{66}$ (note that different longitudinal and transverse modes are reported in this second study) at the lower (superconducting) \Tc, and a kink only in $c_{11}$ at $T_{\rm c}^{\rm Z2}$ \cite{halcrowUltrasoundEvidenceMulticomponent2025}. They suggest that the newer determination of the jump in $c_{66}$ is more reliable than their previous determination of a jump in $(c_{11}-c_{12})/2$ due to an improved background subtraction procedure (taking data in a magnetic field to suppress superconductivity instead of linear extrapolation for the background). 

The authors note that there is no single theory consistent with all of the above observations. In conjunction with other data, they suggest that an $s+id_{xy}$ state provides the best explanation for their data: this state is consistent with a jump in $c_{66}$. However, the previously-reported jump in $(c_{11}-c_{12})/2$ is not consistent with this state---it would require either a $s+id_{x^2-y^2}$ state, or a $d_{xz}+id_{yz}$ state. They also note that symmetry breaking from disorder, and the intrinsically large strain susceptibility of the iron pnictide superconductors \cite{chuDivergentNematicSusceptibility2012}, may complicate the interpretation.

Iron arsenides are particularly challenging for pulse echo ultrasound, as the samples are quite thin and tend to be ``flaky'' along the $c$ axis. The authors of Halcrow \textit{et al.}  \cite{halcrowUltrasoundEvidenceMulticomponent2025} suggest that follow up studies on larger crystals with better alignment should be pursued. It would also be worth attempting to fabricate thin-film ZnO transducers \cite{theussSinglecomponentSuperconductivityUTe22024a}, which can operate with a much smaller transducer area by virtue of working at higher frequencies, avoiding the need for larger, and therefore potentially more disordered, crystals.

\section{FUTURE DIRECTIONS}
\label{se:future}

The case for bulk chiral superconductivity is currently at best weak, at worst negative---we do not have any convincing examples of bulk chiral superconductors. The current best candidates are \ute under pressure, or possibly still \upt, but these are still under investigation. There are, however, several promising new directions for the future.

\subsection{Hexagonal Point Groups}

If the end goal is Majorana zero modes, then spin-triplet superconductivity is generally necessary, otherwise there will always be a pair of Majoranas that will hybridize. If one relaxes this constraint, however, then the singlet $d_{xy}+id_{x^2-y^2}$ state is a fully gapped (in 2D), chiral, and potentially (depending on the band structure) topological superconductor. Such a state is currently being pursued in two dimensions \cite{zhaoTimereversalSymmetryBreaking2023}, where it was suggested that two layers of the known-$d_{x^2-y^2}$ superconductor Bi$_2$Sr$_2$CaCu$_2$O$_{8+x}$, twisted $45^{\circ}$ with respect to one another, could couple to produce a chiral $d_{xy}+id_{x^2-y^2}$ state \cite{canHightemperatureTopologicalSuperconductivity2021}. Although it lacks a Majorana zero mode, this topological state may still be of interest for qubit applications because it is robustness against charge noise \cite{broscoSuperconductingQubitBased2024}.

The natural host for a $d_{xy}+id_{x^2-y^2}$ state in bulk materials is in a hexagonal or trigonal point group: there, the $d_{xy}$ and $d_{x^2-y^2}$ order parameters are naturally degenerate, forming a two-dimensional irrep. (e.g. $E_{2g}$ in $D_{\rm 6h}$.)\footnote{$d_{xz}$ and $d_{yz}$ are degenerate both in 6 fold and also 4-fold point groups, but contain a horizontal line node at $k_z = 0$ therefore their chiral combination does not form a strong topological superconducting state.} While $p_x$ and $p_y$ are also degenerate in these groups, it may be that triplet superconductivity is much harder to find in nature than singlet superconductivity.

A fruitful line of research could be to systematically search for a jump at \Tc in the $c_{66}$ elastic modulus in all hexagonal and trigonal superconductors---the $d_{xy}\pm id_{x^2-y^2}$, $p_x\pm ip_y$, and $d_{xz}\pm id_{yz}$ states all have bilinears that transform in the same way as $\epsilon_{xy}$. These include the recently-discovered AV$_3$Sb$_5$ (A = K, Rb, Cs) \cite{ortiz$mathrmCsmathrmV_3mathrmSb_5$$mathbbZ_2$Topological2020} and TRu$_3$Si$_2$ (T=Y,La) \cite{kuSuperconductingMagneticProperties1980} kagome families of materials\footnote{It should be noted that the presence of a high-temperature (compared to \Tc) CDW transition in both families complicates the symmetry analysis.}; the various incarnations of the superconducting transition metal dichalcogenides (e.g. 2H-NbSe$_2$) \cite{wilsonTransitionMetalDichalcogenides1969}; the C14 laves phases (e.g. CeRu$_2$) \cite{matthiasSuperconductingAlkalineEarth1957}; and various other superconductors of the right symmetry that can be identified via databases such as 3DSC \cite{sommer3DSCDatasetSuperconductors2023}. Of course, care should be taken to choose materials judiciously: for example, MgB$_2$ has the right crystal structure, but is a well-established $s$-wave superconductor.

\subsection{Other experiments that yield symmetry-sensitive information in superconductors}

There are other thermodynamic coefficients that are sensitive to the additional degrees of freedom present in multi-component superconductors. Both the thermal expansion coefficient, $\alpha = (\partial \varepsilon/\partial T)_{\sigma}$, and the elastocaloric coefficient, $\Gamma = (\partial T/\partial \sigma)_S$, are second derivatives of the free energy and thus show jumps at 2$^{\rm nd}$ order phase transitions. These jumps, however, have been studied much less---both theoretically and experimentally---than those of the elastic moduli. Some of the Ehrenfest relations between the elastocaloric coefficients and elastic moduli were worked out in Jerzembeck \textit{et al.} \cite{jerzembeck$T_c$ElastocaloricEffect2024}, and some of the relations for the thermal expansion coefficients were worked out in Mattoni \textit{et al.} \cite{mattoniDirectEvidenceAbsence2025}. Jumps in the elastocaloric coefficient \textit{have} been used as evidence for a transition to a $s+id_{xy}$ state \cite{ghoshElastocaloricEvidenceMulticomponent2025}. However, this was in the $\varepsilon_{xx}$ channel, which is expected to have a jump at any second-order phase transition, and thus did not provide direct evidence for which point group symmetry was broken at the phase transition. 

A cursory analysis suggests that there are no surprises---shear elastocaloric and thermal expansion coefficients have only kinks at \Tc for single-component order parameters, but have jumps for multi-component order parameters\footnote{In principle, measurements of shear thermal expansion coefficients could be even more revealing than those of elastic moduli because which shear coefficients show jumps will depend on the specific ordered state. For example, the $p_x$ state generates a spontaneous $\epsilon_{xx}-\epsilon_{yy}$ strain below \Tc, whereas it does not generate an $\epsilon_{xy}$ strain. However, it is not clear how one could directly measure shear expansion coefficients using traditional methods like dilatometry---perhaps extremely high resolution X-ray scattering experiments could probe this physics.}. It would be highly desirable for such measurements to be pursued further---beyond \sro \cite{jerzembeck$T_c$ElastocaloricEffect2024,mattoniDirectEvidenceAbsence2025}---as the systematic errors in thermal expansion and elastocaloric measurements are entirely different from those encountered in ultrasound. 

\section{CONCLUSION}

Chiral topological superconductivity remains one of the most compelling phases of quantum matter. Despite decades of intense experimental and theoretical effort, no material has as yet provided unambiguous, bulk thermodynamic evidence for a topologically nontrivial superconducting ground state. I have focused on an experimental technique that addresses one specific bottleneck in this search: the lack of experimental probes capable of distinguishing multi-component from single-component superconductivity.

Ultrasound measurements of elastic moduli are uniquely powerful in this context. They are bulk, thermodynamic probes; they are symmetry-resolved; and they directly couple to new degrees of freedom that emerge only in multi-component order parameters. In particular, discontinuities in shear elastic moduli at the superconducting transition constitute direct, detail-independent evidence for a multi-component superconducting state---a necessary (though not sufficient) condition for chiral topological superconductivity in two dimensions.

This program has yielded mixed and often puzzling results across an array of materials. In \sro, independent ultrasound experiments report shear-modulus anomalies consistent with a two-component order parameter \cite{ghoshThermodynamicEvidenceTwocomponent2021,benhabibUltrasoundEvidenceTwocomponent2021}, yet the implied linear strain couplings and transition splittings have not been observed \cite{jerzembeck$T_c$ElastocaloricEffect2024}. In \ute, ambient-pressure superconductivity appears single-component \cite{theussSinglecomponentSuperconductivityUTe22024a}, but under pressure the phase diagram reveals a regime of two interacting order parameters whose symmetry and topology remain unresolved \cite{kamatThermodynamicDiscoveryTetracriticality2026}. \upt, long regarded as a paradigmatic multi-component superconductor \cite{saulsOrderParameterSuperconducting1994}, lacks the expected shear jump but also lacks high-resolution data \cite{thalmeierElasticConstantAnomalies1991}, and recent experiments have raised other issues of sample quality \cite{gannonSpinSusceptibilityTopological2017}. In iron pnictides, shear-modulus anomalies have been reported in two different experiments, but which anomalies (jumps or kinks) and at which transitions is inconsistent between different experiments \cite{grinenkoStateSpontaneouslyBroken2021,halcrowUltrasoundEvidenceMulticomponent2025}.

These results are sobering: either bulk chiral superconductivity is genuinely rare, or material complications such as disorder (of many different varieties) obscure what should be its clear signatures. Ultrasound has not yet delivered a definitive candidate chiral topological superconductor, but it \textit{has} systematically eliminated some candidates.

The path forward is clear: materials with intrinsically degenerate irreps.---particularly those in hexagonal and trigonal point groups---offer the best opportunities. Equally important is the development of complementary thermodynamic probes with comparable symmetry sensitivity, such as elastocaloric and thermal expansion measurements (and particularly those sensitive to shear strains). Most crucially, future work must confront disorder, inhomogeneity, and mesoscale structure as unavoidable features of real materials. If chiral superconductivity exists, it will likely reveal itself only through a coordinated, symmetry-based experimental effort of the kind outlined here.

\begin{summary}[SUMMARY POINTS]
\begin{enumerate}
\item Chiral topological superconductivity in two dimensions requires a multi-component superconducting order parameter.

\item Ultrasound measurements of elastic moduli provide a bulk, thermodynamic, and symmetry-sensitive probe that is ideally suited for detecting multi-component order parameters.

\item In \sro, elastic moduli measurements suggest a multi-component order parameter, but the implied linear coupling of \Tc to shear strain has not been observed, and other experiments rule out a topological state.

\item In \ute, elastic measurements rule out a multi-component state at ambient pressure, but a multi-component state under pressure is still a possibility, and elastic measurements have clarified where in the phase diagram this might occur. 

\item In \upt, the expected shear modulus jump was not observed and requires reinvestigation with a higher-resolution, modern experiment. 

\item Quantitative inconsistencies and strong sensitivity to sample quality highlight the roles of disorder and inhomogeneity in candidate materials, and suggest the need for theory that incorporates the effects of disorder beyond the effective medium approximation.

\item The lack of confirmed chiral superconductors underscores the need for systematic, symmetry-guided searches rather than case-by-case investigations.
\end{enumerate}
\end{summary}

\begin{issues}[FUTURE ISSUES]
\begin{enumerate}
\item Are there materials with hexagonal or trigonal symmetry that host a naturally degenerate, singlet or triplet superconducting order parameter leading to robust chiral superconductivity?

\item Can high-resolution shear-modulus measurements be performed systematically across large materials databases to rule in or out entire classes of superconductors?

\item What role do disorder, stacking faults, and mesoscale phase separation play in generating apparent multi-component signatures in intrinsically single-component superconductors?

\item Do the interacting superconducting phases in \ute form a time-reversal-symmetry-breaking state, and if so, what is its topology?

\item Can elastocaloric and thermal expansion measurements be developed into symmetry-sensitive thermodynamic probes comparable to ultrasound?

\item To what extent can surface-sensitive probes reporting Majorana-like signatures be reconciled with bulk thermodynamic constraints?

\item Is chiral superconductivity intrinsically rare, or have its most favorable material realizations simply not yet been identified?
\end{enumerate}
\end{issues}

\section*{DISCLOSURE STATEMENT}
The authors are not aware of any affiliations, memberships, funding, or financial holdings that
might be perceived as affecting the objectivity of this review.

\section*{ACKNOWLEDGMENTS}
B. J. R. acknowledges helpful discussions and editing provided by Nico Huber, Sahas Kamat, Andrew Mackenzie, Steve Kivelson, Sri Raghu, and Yoshiteru Maeno.

%

\bibliographystyle{ar-style4}
\bibliography{references}

\end{document}